\documentclass[sigconf,screen]{acmart}
\AtBeginDocument{%
  }

\newcommand{\MName}{EffiHolmes}

\usepackage{xurl}  % ⭐ 推荐这个，最省事
\usepackage{comment}
\usepackage{pgfplots}
\usepgfplotslibrary{groupplots}
\usepackage{amsmath}
\usepackage{graphicx}
\usepackage{enumitem}
\usepackage{hyperref}
\usepackage{mdframed}
\usepackage{tcolorbox}
\usepackage{multirow}
\usepackage{tikz}
\usepackage{xurl}
\usepackage{microtype}
\usepackage{tikz-qtree}
\usepackage{pgfplots}
\usepackage{siunitx}
\usepackage{enumitem}
\usepackage{booktabs}
\usepackage{graphicx}
\usepackage{booktabs}   % 对应图片中的三线表风格
\usepackage{multirow}   % 用于合并行
\usepackage{graphicx}   % 用于 resizebox
\usepackage{colortbl}   % 用于背景上色
\usepackage[linesnumbered, ruled, vlined]{algorithm2e}
\setlist[itemize]{leftmargin=*}
\definecolor{lightgray}{rgb}{0.92, 0.92, 0.92}
\usepackage{balance}
\pgfplotsset{compat=1.8}
\usepgfplotslibrary{statistics}

\definecolor{modelA}{RGB}{245,250,249}  % 几乎白，只比背景深一点点
\definecolor{modelB}{RGB}{235,244,242}  % 你现在的原 modelA
\definecolor{modelC}{RGB}{215,235,230}  % 你现在的原 modelB

\definecolor{morandi1}{RGB}{55,85,75}
\definecolor{morandi2}{RGB}{85,115,100}
\definecolor{morandi3}{RGB}{115,145,130}

\definecolor{deepred}{RGB}{140,40,40}

\usepackage{colortbl}
\usepackage{pgf} % for \pgfmathsetmacro

\definecolor{modelA}{RGB}{245,250,249}
\definecolor{modelB}{RGB}{235,244,242}
\definecolor{modelC}{RGB}{215,235,230}

\definecolor{heatLow}{RGB}{245,250,249}   % light
\definecolor{heatHigh}{RGB}{195,225,218}  % dark

\setcopyright{cc}
\setcctype{by}
\acmDOI{10.1145/3832783.3834353}
\acmYear{2026}
\copyrightyear{2026}
\acmISBN{979-8-4007-2882-2/2026/10}
\acmConference[ASE '26]{Proceedings of the 41st IEEE/ACM International Conference on Automated Software Engineering}{October 12--16, 2026}{Munich, Germany}
\acmBooktitle{Proceedings of the 41st IEEE/ACM International Conference on Automated Software Engineering (ASE '26), October 12--16, 2026, Munich, Germany}
\acmSubmissionID{ase26main-p417-p}
\received{2026-03-26}
\received[accepted]{2026-06-18}

\begin{document}

% A short title is provided for the running head.
\title[\MName: Differential Profiling-Guided Repository Level...]
{\MName: Differential Profiling-Guided Repository Level Time Inefficiency Fix Localization}

% =========================================================
% Authors
% Each author must be declared separately.
% =========================================================

\author{Haowen Yang}
\orcid{0009-0008-7709-2923}
\affiliation{%
  \institution{Hong Kong University of Science and Technology (Guangzhou)}
  \city{Guangzhou}
  \country{China}
}
\email{hyang464@connect.hkust-gz.edu.cn}

\author{Yun Peng}
\orcid{0000-0003-1936-5598}
\affiliation{%
  \institution{Chinese University of Hong Kong}
  \city{Hong Kong}
  \country{Hong Kong}
}
\email{ypeng@cse.cuhk.edu.hk}

\author{Zishuo Ding}
\authornote{Corresponding author.}
\orcid{0000-0002-0803-5609}
\affiliation{%
  \institution{Hong Kong University of Science and Technology (Guangzhou)}
  \city{Guangzhou}
  \country{China}
}
\email{zishuoding@hkust-gz.edu.cn}

% Three complete names should normally fit in the running head.
% Only enable this if the complete author list does not fit:
%
% \renewcommand{\shortauthors}{H. Yang, Y. Peng, and Z. Ding}

% =========================================================
% Abstract
% Sections/001_Abstract.tex should contain only abstract text,
% without \begin{abstract} and \end{abstract}.
% =========================================================
\begin{abstract}

\label{sec:abstract}

Large software systems often suffer from time inefficiencies, leading to excessive execution time despite functional correctness.  
Localizing the fix locations of such inefficiency issues is notoriously difficult.
% Existing work on code inefficiency optimization falls short as they primarily target the function level, failing to capture the complex, cross-module interactions in the repository. \py{There is a gap here. Localizing the fix -> code opt -> fault localization. You need a more smooth connection between these concepts.}
Applying existing fault localization paradigms to this task presents significant challenges.
Unlike functional bugs, time inefficiencies do not cause crashes or test failures. 
They therefore provide neither binary oracles nor stack-trace localization clues, making traditional fault localization and recent LLM-based methods not applicable.
While runtime profiling provides alternative clues, it faces some challenges in repository-level settings.
Single-run profiling is non-discriminative, failing to distinguish inefficiency hotspots (i.e., time-consuming functions) from execution noise.
Furthermore, existing profilers struggle to extract the execution paths relevant to inefficiency from the vast background execution.
Crucially, even with the extracted execution paths, a semantic gap still persists between the observed inefficiency and where to fix.

% 但是对于 repo level 的 time inefficiency 问题来说，1. aggregated profilers 输出的是聚合计时数据 会丢失 调用链这种关键的信息；2. 非aggregated profilers 能保留调用链信息，但是由于 repo level ，trace 会很大， too much profiling information with a lot of noise. 会让 llm loss
% However, for repository-level time inefficiency issues, existing profiling information are either too coarse to capture execution structure or too noisy to support precise localization.
% In addition, static single-run profiling are often non-discriminative, as true inefficiency signals are submerged in execution noise.

In this paper, we propose \textsc{EffiHolmes}, an LLM-based framework for localizing fix locations of repository-level time inefficiency issues.
First, it employs differential profiling (default vs.\ scaled workloads) to pinpoint inefficiency hotspots. 
Second, to extract the execution paths relevant to inefficiency, it performs execution path extraction, constructing compact execution paths that connect hotspots to the reported inefficient function. 
Third, to bridge the gap between the hotspots and fix locations, it employs guided LLM reasoning with domain heuristics to locate the inefficiency logic.
To facilitate the evaluation, we introduce \textsc{RepoEffi-Bench}, the first benchmark for repository-level inefficiency localization, consisting of 140 high-quality time inefficiency issues collected from popular Python repositories. 
The results show that \textsc{EffiHolmes} consistently outperforms state-of-the-art retrieval-based, agent-based, and profiling-based baselines, with gains of 4.29 pp on GPT-5.1 file-level Acc@3 and 15.00 pp on qwen3-4b function-level Acc@5 over the best baseline.
It also remains robust across model capacities.

\begin{figure}[t]
    \centering
    \includegraphics[width=\columnwidth]{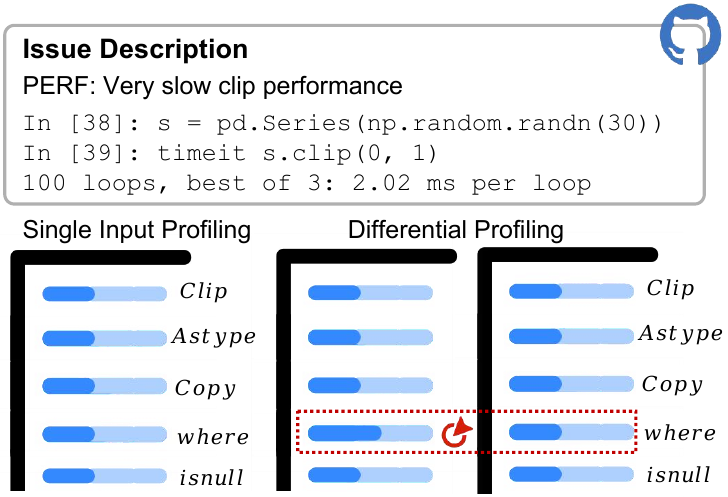}
    \caption{Single-Run Profiling vs Differential Profiling.}
    \label{fig:Intro_Challenge3}
\end{figure}

\end{abstract}

% =========================================================
% CCS Concepts
% IMPORTANT:
% Replace this entire part with the exact code generated by
% the ACM CCS tool. The XML and \ccsdesc must describe the
% same concept.
% =========================================================


% Paste the matching \ccsdesc commands generated by the tool.
% For example:
% \ccsdesc[500]{Software and its engineering~Software performance}

% =========================================================
% Keywords are mandatory.
% =========================================================

\maketitle

% =========================================================
% Main content
% =========================================================

\section{Introduction}
\label{sec:intro}

Large software systems often suffer from time inefficiency issues~\cite{zhao2022large}, which cause excessive execution time despite the program remaining functionally correct~\cite{zhao2020performance}.
Such issues can lead to severe consequences, such as budget overrun~\cite{zaman2012qualitative}, project delay \cite{yang2018not, zhao2022large}, system failures~\cite{SpotifyCrash}, and even financial losses \cite{yang2025towards, chen2023toward, Pokeman}. 
For instance, a latency issue~\cite{Freqtrade_stop_loss} in the popular trading bot \textit{Freqtrade}~\cite{Freqtrade} resulted in additional processing time when the bot attempted to automatically sell assets to protect users during market crashes. 
Due to this delay, the system failed to respond immediately to price drops, causing significant financial losses. 

Accurate localization of fixes (i.e., identifying where code changes should be applied to reduce execution time) is a fundamental task to addressing these time inefficiency issues~\cite{meng2024empirical}.
However, this task is often difficult in practice, as developers must manually construct input workloads to trigger the inefficiency, trace long and complex call chains across multiple modules, and reason about where the unjustified overhead is introduced~\cite{baltes2015navigate}.
For example, in \textit{Xarray}~\cite{xarray} Issue \#9692~\cite{xarray_9692}, developers faced a 30x slowdown in the \texttt{DataArray.quantile} function.
It took four experts twelve days to determine where to fix, as it required tracing a deep call chain from the API to low-level thread management and reasoning across \textit{Xarray}’s wrapper layers and backend.

Despite the severe impact of time inefficiencies and the significant manual effort required for diagnosis, current approaches remain inadequate for addressing these issues.
The task of pinpointing specific code regions responsible for inefficiency issues falls under the domain of fault localization (FL)~\cite{gazzola2018automatic}.
Yet, applying existing FL paradigms to this task presents significant challenges.
Traditional FL techniques are largely inapplicable to this task due to their reliance on explicit correctness signals. 
Whether employing \textit{Spectrum-based (SBFL)}~\cite{jones2005tarantula,abreu2007ochiai, abreu2009spectrum, raselimo2019spectrumCFG, reis2019demystifying, wong2014dstar, zhang2011localizing}, \textit{Mutation-based (MBFL)}~\cite{papadakis2013metallaxis,moon2014muse}, or \textit{Machine Learning-based (MLFL)}~\cite{li2021fault, li2019deepfl, lou2021boosting} strategies, \textbf{these methods share a fundamental dependency on a binary oracle to distinguish failing executions from passing ones. }
This assumption fails for inefficiency issues, which manifest as silent performance degradation rather than explicit failures~\cite{jin2012understanding, han2016empirical}.
More recently, researchers have turned to LLM-based FL approaches that exploit the semantics of issue reports rather than correctness signals.
Whether employing hierarchical reasoning~\cite{xia2024agentless, chang2025bridging, ouyang2024repograph, jiang2025cosil} or agent-based navigation~\cite{chen2025locagent, yang2024sweagent, yu2025orcaloca, wang2024openhands, aorwall_moatless_tools_2025}, these methods share a critical limitation: \textbf{they rely heavily on explicit localization clues within the issue description}, such as stack traces or error messages. These clues are critical for guiding models to narrow the search space in complex repositories.
However, time inefficiency issues differ from functional ones as they do not produce explicit failure artifacts, such as stack traces, that can serve as clues for localization.
As a result, the assumption underlying LLM-based FL does not hold in repository-level inefficiency issues.

In the absence of these clues, we use runtime profiling as an alternative source.
However, this introduces new challenges.
% challenge 1
\textbf{First, static single-run profiling fails to distinguish inefficiency signals from execution noise.} 
A single run provides only a static snapshot where performance hotspots are often submerged in noise. 
As illustrated in Fig.\ref{fig:Intro_Challenge3} using \textit{Pandas} Issue \#15400, the faulty function executes in merely 0.002 seconds. 
At this low magnitude, the inefficiency is numerically indistinguishable from noise such as \texttt{astype} or \texttt{copy}. 
The hotspot, \texttt{where}, is drowned out because standard inputs are too small to allow the inefficiency to dominate. 
It only emerges when the input is scaled, as shown in Fig.\ref{fig:Intro_Challenge3}: the execution-time hotspot explodes while the noise remains stable. 
% challenge 2
\textbf{Second, existing profilers struggle to extract the execution paths relevant to inefficiency from the vast background execution.}
Aggregated profilers (e.g., cProfile~\cite{python_cprofile}) provide insufficient information.
They highlight which functions consume time but flatten the execution history.
By discarding calling relationships, they show the symptom but hide the path between upstream control logic and downstream hotspots. 
In contrast, raw trace profilers (e.g., VizTracer~\cite{viztracer}) contain excessive noise.
While they preserve the full execution history, they generate millions of low-level events.
This results in an extremely low signal-to-noise ratio and buries the execution paths relevant to the inefficiency.
% Without specific domain knowledge to filter this raw data, LLMs do not know how to analyze the raw information and pinpoint the fix location.
% challenge 3
\textbf{Third, profiler-reported hotspots do not reliably indicate fix locations.}
Prior work~\cite{zhao2020performance} shows that profiler-reported hotspots primarily indicate where execution time accumulates, rather than where the issues originate.
Execution hotspots can be amplified by repeated invocations or expensive execution paths, while the control logic that triggers such behavior may reside elsewhere in the code~\cite{weber2021white}.

%%% figure
\begin{figure*}[t]
    \centering
    \includegraphics[width=1\textwidth]{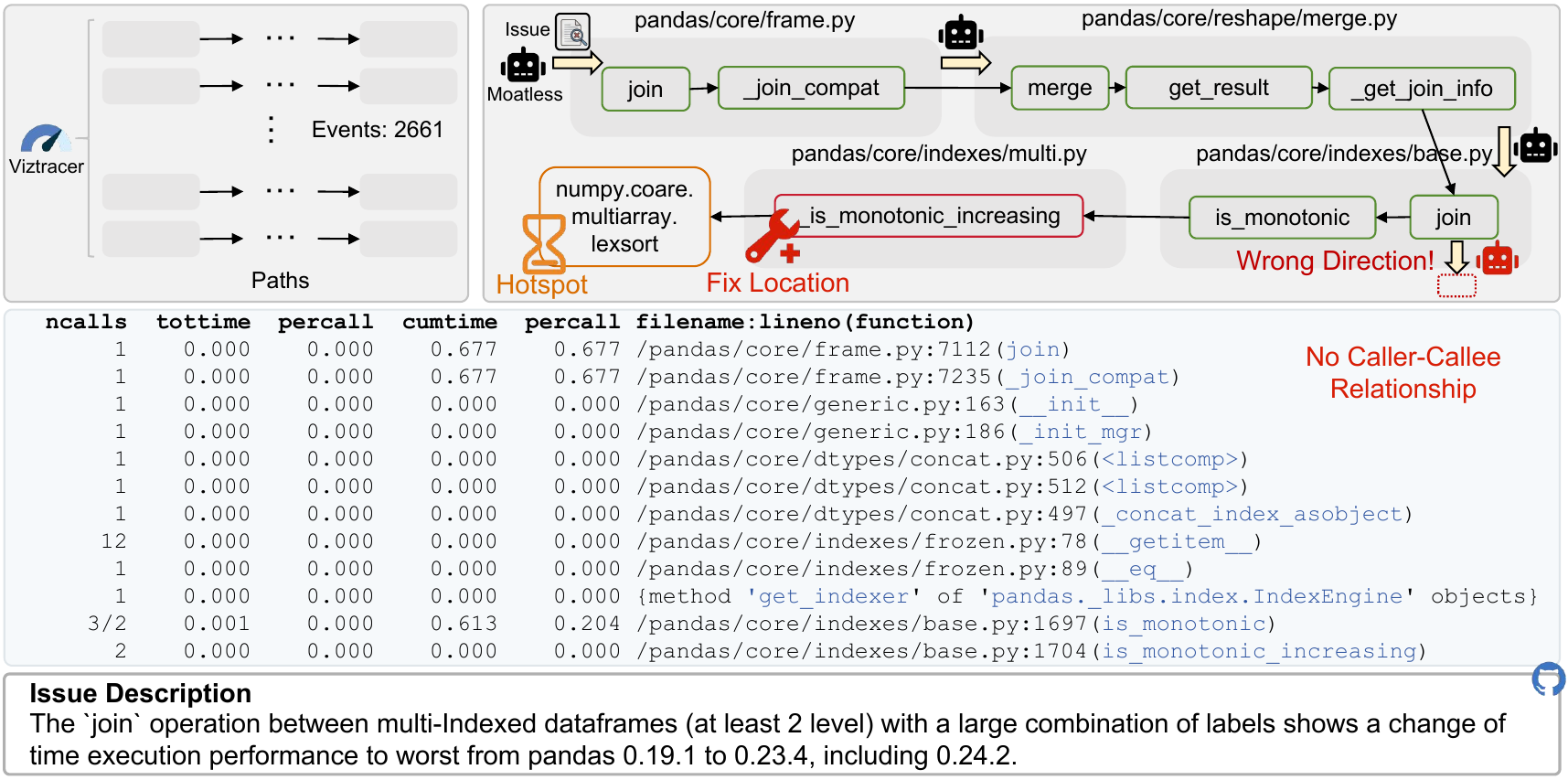}
    \caption{Motivating Example of \textsc{EffiHolmes}.}
    \label{fig:motivating_example}
\end{figure*}

To solve these challenges, we propose \textsc{EffiHolmes}, an LLM-based framework that localizes where to fix time inefficiencies.
% by combining differential profiling with structure-aware trace distillation.
Specifically, it comprises three main components.
% 1) Differential Trace Generation. 
\textbf{To distinguish the inefficiency signal from execution noise}, we construct a differential test pair (default vs.\ scaled execution) and generate time-annotated execution traces, which allows inefficiency hotspots to be exposed through comparison.
% stage2 constructs execution evidence at a suitable granularity 并在其中确定了 hotspots 和 fix location 的关系.
% 2) Execution Paths Distillation. 
\textbf{To extract the execution paths relevant to inefficiency}, we 
% quantify differential execution costs to pinpoint hotspots and 
distilled the execution paths connecting these hotspots to the reported inefficient function from the raw execution traces.
Then we compress the resulting paths by grouping those with identical call-chain structures.
Within each group, the critical execution path with the highest cumulative differential time is retained.
% stage 3 
% 3) Guided LLM Reasoning.
\textbf{To bridge the semantic gap between profiler-reported hotspots and actual fix locations}, we employ domain-guided reasoning. 
Even with extracted critical execution paths, LLMs often pinpoint where time accumulates rather than where the issue originates. 
Therefore, this stage synthesizes structured runtime evidence with domain reasoning heuristics, explicitly guiding the LLM to trace upstream to the defective control logic.

% We evaluate \textsc{EffiHolmes} on \textit{RepoEffi-Bench}.
Additionally, to evaluate our approach, we introduce \textit{RepoEffi-Bench}, a new benchmark specifically curated for repository-level time inefficiency issue fix localization. 
Existing benchmarks are limited by their narrow function-level scope~\cite{huang2024effibench, du2024mercury, ye2025llm4effi, peng2025coffe}, reliance on unrealistic data~\cite{he2025swe}, or leak the fix location in the description~\cite{chen2025locagent}.
\textsc{RepoEffi-Bench} addresses these gaps by validating real-world, developer-reported issues to ensure a practical and leak-free evaluation.
Experimental results demonstrate that \textsc{EffiHolmes} substantially outperforms state-of-the-art (SOTA) retrieval-, agent-, procedure- and profiling-based baselines, achieving gains of 4.29 pp on GPT-5.1 file-level Acc@3 and 15.00 pp on qwen3-4b function-level Acc@5 over the best baselines.
Beyond peak performance, results also reveal that \textsc{EffiHolmes} remains robust across model scales at higher localization cutoffs, enabling a lightweight 4B model to reach 83.57\% \texttt{File-Acc@5} and 70.71\% \texttt{Func-Acc@5}, comparable to baselines running on models with \texttt{8×} as many parameters.

%%% Contribution
We summarize our contributions as follows.
\vspace{-0.5em}
\begin{itemize}

    \item We propose \textsc{EffiHolmes}, the first
    LLM-based framework that enables repository-level time inefficiency fix localization.
    It isolates inefficiency signals from noise, reconstruct inefficiency relevant contexts, and accurately pinpoint fix locations in repository-level code.

    \item We curate \textsc{RepoEffi-Bench}, a new benchmark specifically designed for real-world repository-level inefficiency localization.
    Unlike existing benchmarks that focus primarily on function-level inefficiency issues or leak fix location. RepoEffi-Bench comprises repository-level, leakage-free inefficiency cases raised and confirmed by developers.

    \item Extensive experiments demonstrate the effectiveness of \textsc{EffiHolmes} in localizing inefficiency fixes, consistently outperforming baselines across multiple LLM scales.
\end{itemize}

\vspace{-1em}
% \py{The introduction is too long, try to shorten it if there is no enough space for other sections.}

\section{Motivation}
\label{sec:mot_and_background}

%%% Version 1
% \subsection{Motivating Example}
Fig.~\ref{fig:motivating_example} shows a performance regression issue from \textit{pandas}: a join operation on \texttt{DataFrames} with a \texttt{MultiIndex} became significantly slower than in previous versions.
We inspected the results of SOTA agent-based issue localization methods~\cite{aorwall_moatless_tools_2025,jiang2025cosil,chen2025locagent} and standard profiling techniques.
None successfully identified the correct fix location.
We analyze the failure modes of these approaches below.

\textbf{Failure of Agent-Based Methods.}
We traced the trajectory of SOTA agent-based method
% Moatless~\cite{aorwall_moatless_tools_2025} 
on this issue.
As shown in Fig.~\ref{fig:motivating_example}, starting from the issue description, the method successfully navigates to \texttt{Index.join} but fails to pinpoint the root cause within the function.
Inside \texttt{Index.join}, the agent encountered the property check: \texttt{if self.is\_monotonic\:}.
Lacking runtime feedback, the agent relied on semantic intuition, mistakenly treating this computed property as a trivial field lookup ($O(1)$).
It failed to recognize that accessing \texttt{self.is\_monotonic} triggers a heavy computation.
Consequently, it excluded the property's implementation (\texttt{pandas/core/indexes/multi.py}) from its search scope, thereby completely overlooking the fix location.

\begin{figure*}[t]
    \centering
    \includegraphics[width=1\textwidth]{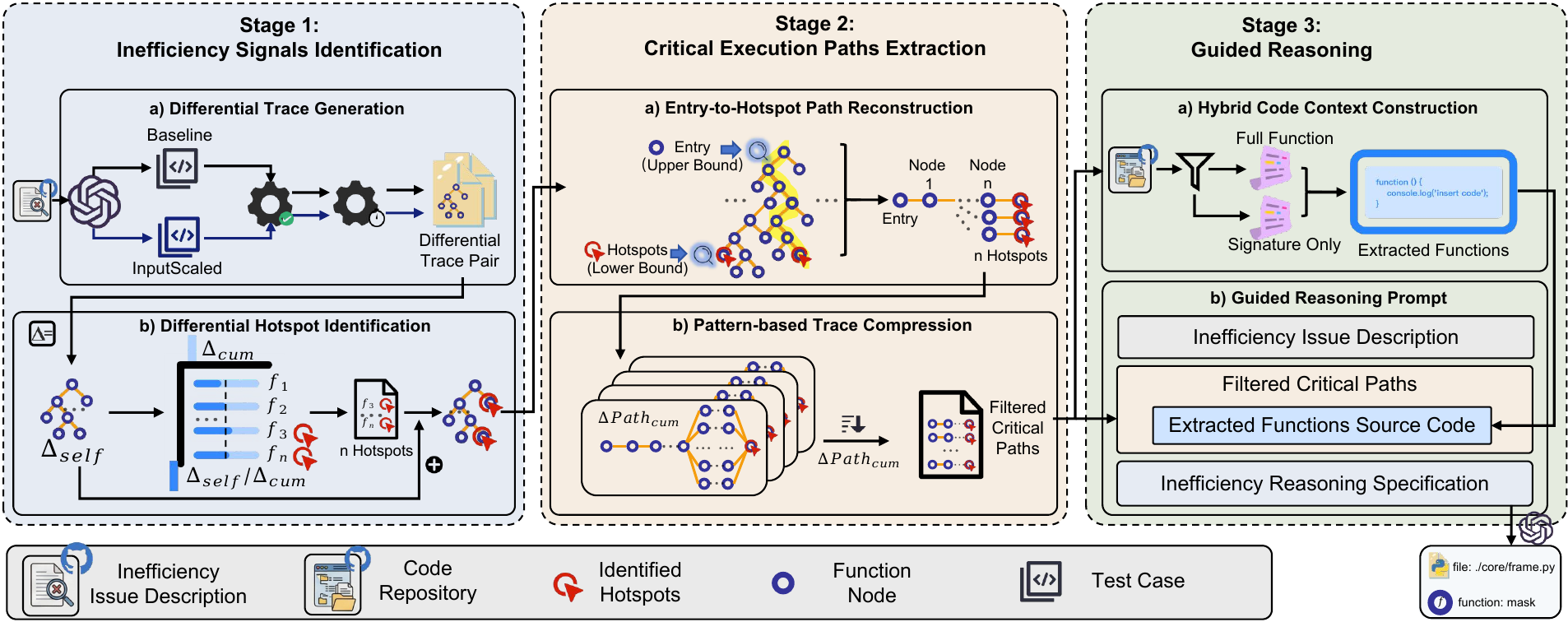}
    \caption{Overview of \textsc{EffiHolmes}. The framework takes a repository and an issue description as input. It generates a differential trace pair to identify hotspots, extracts and compresses critical execution paths, and guides the LLM to locate the fix location.
    }
    \label{fig:overview}
\end{figure*}

\textbf{Failure of Existing Profiling Tools.}
Our analysis also confirms that standard profiling results are insufficient, due to three key observations:
\noindent\textbf{(1) Single-input profiling fails to reveal complexity.}
Standard profilers report absolute execution time for a single run. 
Without comparing execution scaling across different input sizes, the complexity is indistinguishable from other operations.
\noindent\textbf{(2) Aggregation obscures causal chains while tracing drowns logic.}
Existing profilers fail to isolate the structural root cause from the background noise.
\textit{Statistical profilers} (e.g., cProfile) flatten the execution hierarchy.
As shown in the \textbf{bottom table of Fig.~\ref{fig:motivating_example}}, the profile lists functions with aggregated statistics (e.g. \texttt{is\_monotonic}, 0.613s cumtime, 3/2 calls) but explicitly lacks caller-callee relationships.
This flattened view prevents developers from identifying which specific invocation in the recursive stack triggered the latency.
Conversely, \textit{trace profilers} (e.g., VizTracer) preserve temporal details but suffer from structural obscuration.
As illustrated in the \textbf{top-left of Fig.~\ref{fig:motivating_example}}, the tool records over 2,600 events in a mere 0.6s window.
High-level algorithmic logic is heavily interleaved with low-level implementation details, creating a complex web of events that makes it cognitively demanding to reconstruct the primary inefficiency flow.
\noindent\textbf{(3) Profiler-reported hotspots do not reliably indicate fix locations.}
Most critically, there is a gap between the symptom and the cause, as shown in the \textbf{top-right schema of Fig.~\ref{fig:motivating_example}}.
The profiler flags the primitive \texttt{numpy.lexsort} as the \textbf{Hotspot} (marked in orange).
However, the \textbf{Fix Location} (marked in red) actually lies upstream in \texttt{pandas/core/indexes/multi.py}, within the \texttt{is\_monotonic\_increasing} property.

From Observation 1, we find that static or single-input profiling fails to distinguish intrinsic algorithmic complexity from normal linear execution costs. This motivates us to isolate the true inefficiency signal by constructing a differential test pair (baseline vs. scaled execution) and generating time-annotated traces, which allows algorithmic hotspots to be clearly exposed through differential comparison. 
From Observation 2, we find that raw traces entangle critical logic with vast low-level implementation details, obscuring the structural root cause. This motivates us to disentangle and compress the inefficiency relevant execution paths. We distill specific execution paths connecting hotspots to the entry point and group them by call-chain structure, retaining only the representative path with the highest cumulative differential time to filter out noise. 
From Observation 3, we find that a significant semantic gap exists between the symptom (hotspot) and the cure (fix location), often causing LLMs to fixate on leaf computations where time accumulates. 
To bridge this gap, we integrate domain-guided reasoning constraints. 
Instead of blindly following hotspots, we explicitly guide the LLM to prioritize upstream control-layer logic over leaf computations and validate candidates against structural evidence, ensuring the search converges on the defective logical decision.

\section{The \textsc{EffiHolmes} Framework}
\label{sec:methodology}

\subsection{Overview}

In this section, we introduce \textsc{EffiHolmes}, an LLM-based framework for localizing time inefficiency fixes.
As shown in Figure~\ref{fig:overview}, \textsc{EffiHolmes} takes an inefficiency issue report and the corresponding repository as inputs and returns the identified fix location of the reported inefficiency.
It consists of three main stages: inefficiency signals identification, critical execution paths extraction, and guided LLM reasoning. 
Inefficiency signals here refer to profiler-generated traces and hotspots.
To isolate the differential hotspots from execution noise, \textsc{EffiHolmes} performs differential profiling by constructing a default and a scaled execution pair.
It then quantifies the differential costs to isolate the hotspots (Section~\ref{sec:diff trace gen}). 
To extract the execution paths relevant to inefficiency, \textsc{EffiHolmes} distills the execution paths linking these hotspots to the issue's entry point (Section~\ref{sec:causal paths distill}).
Additionally, to reduce redundancy and provide concise context, \textsc{EffiHolmes} uses a path-compression mechanism to collapse structurally identical paths into critical representatives. 
To bridge the semantic gap between differential hotspots and actual fix locations, \textsc{EffiHolmes} leverages domain heuristics to guide the LLM in synthesizing dynamic context with static information, thereby pinpointing the final fix location (Section~\ref{sec:reasoning}).

\subsection{Inefficiency Signals Identification}
\label{sec:diff trace gen}

This stage generates the differential traces and pinpoints the inefficiency hotspots that serve as the inputs to subsequent execution path distillation.
Given an inefficiency issue and its corresponding repository, we construct and profile two executions that differ only in workload scale. 
Then, inefficiency hotspots are identified through comparison.

\subsubsection{Differential Trace Generation.}

\textbf{Differential Test Pair Generation.} Constructing a reliable differential test pair is critical for isolating inefficiency signals.
Directly prompting LLMs to generate large-scale inputs often produces syntactically valid but semantically benign workloads that fail to trigger the target inefficiency~\cite{liu2024evaluating}. For instance, in \textit{scikit-learn}~\cite{Scikit-learn} issue \#24840~\cite{Sklearn24840}, simply increasing the row count does not trigger the slowdown; the performance degradation only manifests when the data remains \emph{sparse}, i.e., more than 99\% of the matrix entries are \texttt{NaN}. 
Therefore, we do not scale inputs blindly, but use issue-specific cues as the primary scaling target when available.
When explicit scale cues are unavailable, the LLM infers a scale-only target from the baseline script and the issue report, while preserving the original API sequence, non-target code, input data types and distributions, and the target API call.
Based on either the explicit cue or the inferred target, \textsc{EffiHolmes} constructs the \textit{scaled test case} ($T_{scaled}$), while the \textit{default test case} ($T_{base}$) is kept unchanged at its original scale to capture workload-independent overheads.
To ensure that the generated pair is usable for differential profiling, we validate each $(T_{base}, T_{scaled})$ using an execution-based generation loop. Specifically, we require that both test cases execute successfully without introducing new failures such as crashes, and that profiling can be completed successfully on both executions.
In addition, we require $T_{scaled}$ to incur larger runtime overhead relative to $T_{base}$ under the same profiling setup, so as to exclude workload pairs that are effectively indistinguishable at runtime. 
If these conditions are not met, we treat the generated workload as unusable for differential profiling and regenerate $T_{scaled}$ accordingly, with a bounded retry budget of five rounds. 
Finally, the accepted $(T_{base}, T_{scaled})$ pair forms the \textbf{Differential Test Pair} used for subsequent differential hotspot identification.

\textbf{Trace Profiling.}
To obtain the execution traces required for differential analysis, we employ a non-aggregated profiler to instrument both executions.
Specifically, we use VizTracer~\cite{viztracer} as the profiling backend.
VizTracer preserves the full hierarchical call structure of program execution,
which is essential for reconstructing causal paths from entry points to performance hotspots.
It also records deterministic entry and exit timestamps for each function call,
enabling precise computation of differential execution costs between
$T_{base}$ and $T_{scaled}$.
Finally, its low-overhead C-based implementation minimizes profiling distortion,
ensuring that the observed performance differences reflect true algorithmic behavior under scaled workloads rather than tracing overhead.

\subsubsection{Differential Hotspot Identification}

We identify functions whose runtime cost exhibits significant growth under scaled workloads.
To this end, we quantify execution growth using the \emph{Differential Self-Time}, defined as the difference in self execution time (i.e., the time spent in the function body excluding callees) of a function node between the scaled and default executions.
Formally, for each function node $n$ in the dynamic call tree, we compute:
% \vspace{-1.5em}

\begin{equation}
\Delta_{self}(n) = T_{scaled}(n).\textit{self} - T_{base}(n).\textit{self}.
\end{equation}

To assess the relative contribution of each node to the overall performance regression,
we further define the \emph{Normalized Differential Contribution} (NDC) of node $n$ as:
\begin{equation}
\mathrm{NDC}(n) = \frac{\Delta_{self}(n)}{\sum_{m \in \mathcal{N}} \Delta_{self}(m)},
\end{equation}
where $\mathcal{N}$ denotes the set of all function nodes in the call tree.

We rank function nodes by their NDC values and select the top-$k$ nodes as inefficiency hotspots.
These hotspots form the \emph{lower bound} of the localization search space, corresponding to functions where inefficiencies manifest most prominently.

\subsection{Critical Execution Paths Extraction}
\label{sec:causal paths distill}
Given the differential execution traces obtained from the previous stage, this subsection extracts them into a compact set of critical execution paths.
The goal is to extract the execution paths that preserves the execution dependencies through which the inefficiency propagates while eliminating irrelevant background noise. 
To achieve this, we employ a two-phase process. 
First, we reconstruct the execution path from the entry point to the hotspot, tracing the call chain from the execution entry point to the identified hotspot.
Second, we apply pattern-based trace compression to remove redundancy introduced by repeated dynamic executions (e.g., loop iterations), grouping structurally identical call chains and retaining only a single representative critical execution path (the one with the maximum cumulative differential self-time) from each group.

\subsubsection{Entry-to-Hotspot Path Reconstruction}

Identifying hotspots alone is insufficient for fix localization, as the fix location may reside in an upstream function that introduces inefficient control logic~\cite{zhao2020performance, weber2021white}.
This limitation of profiling-based approaches has been noted in prior work~\cite{jin2012understanding}, which observes that long propagation chains and boundary-level overheads often obscure the fix locations of inefficiency issues.
Here, we use the term \emph{execution paths} in a structural sense,
referring to call-chain dependencies through which inefficient logic propagates during execution, as captured by the profiler generated traces.

We anchor path reconstruction between two bounds: the \emph{upper bound}, defined as the \emph{entry point} function invoked by the reproduction script, and the \emph{lower bound}, defined as the set of identified differential hotspots.
Given the call tree of the scaled execution, \textsc{EffiHolmes} enumerates all execution paths that connect the entry point to each hotspot.
These paths represent concrete propagation chains along which inefficient logic flows from the user-observable symptom to the system-detected performance hotspots.

\subsubsection{Pattern-Based Trace Compression}

Even after path reconstruction, the resulting set of execution paths often contains substantial redundancy.
This redundancy arises because dynamic tracing records every execution instance.
For example, an inefficient loop iterating many times generates numerous execution paths with identical call-chain structure.

To eliminate such redundancy, we apply a pattern-based trace compression strategy.
Each frame is represented by its repository-relative source file and function name, and each execution path is normalized as the ordered sequence of these frame identifiers.
Line numbers are excluded from the structural identifier so that different dynamic occurrences of the same call chain share the same representation.
Two paths are considered \emph{structurally identical} if and only if their normalized frame sequences are identical, and are therefore assigned to the same path pattern.
For each path, we compute its cumulative differential self-time by summing, over all frames, the scaled-execution self-time of the occurrence minus the mean baseline self-time of the corresponding file--line--function identity.
Within each pattern, we retain the path with the maximum cumulative differential self-time; ties are broken in favor of the earliest target occurrence in the scaled trace.
This procedure preserves one deterministic representative carrying the strongest differential signal for each distinct calling structure.

%%% ------------------------------------------------ %%%

\subsection{Guided Reasoning}
\label{sec:reasoning}

This stage guides an LLM to identify the inefficiency fix location by reasoning over the compact set of critical execution paths produced in the previous stage.

\subsubsection{Hybrid Code Context Construction.}
The critical execution paths with timing information localize where the performance degradation manifests, but effective fix localization still requires code-level context to interpret why the inefficiency arises.
However, repository-level code is often too large to be fully injected into the LLM context.
To balance contextual completeness and scalability, we adopt a compact hybrid code injection strategy.
For functions with at most $k$ lines, we retain the full source code.
For longer functions, we use a compressed representation that preserves decorators, the function signature, the docstring, and code lines containing call expressions. 
This design is grounded in previous findings on long-context limitations in LLMs~\cite{rando2025longcodebench,du2025context}.
We retain call-expression lines because they provide lightweight but informative evidence about inter-procedural dependencies and invocation structure, which is crucial for reasoning about how inefficiency propagates along execution paths, while avoiding the context dilution caused by injecting full long-function bodies.
We compare $k\in\{20,60,100\}$ and find that $k{=}100$ performs best overall, although the gain is limited (e.g., Top-1 accuracy improves from 36.43\% at $k{=}20$ to 40.00\% at $k{=}100$).
Given performance-context tradeoff, we use the 100-line setting in the final implementation.

\subsubsection{Guided Reasoning Constraints.}
Given the critical execution paths and the hybrid code context constructed above, we impose explicit reasoning constraints in the prompt to guide the LLM toward fix-location localization rather than hotspots.
The prompt asks the model to return a ranked shortlist of candidate fix locations together with supporting code evidence.
First, the model considers all repository nodes along the extracted entry-to-hotspot paths and distinguishes control-layer functions from other nodes.
Control-layer functions that govern data distribution, batching, iteration structure, algorithm selection, or input scale are treated as high-leverage candidates.
However, this preference is not absolute: a leaf function with dominant differential self-time may rank higher when concrete source-code evidence supports it as the likely fix location.
Second, the model is required to validate each predicted fix location against concrete source-code evidence.
Specifically, it must identify structural indicators of inefficiency, such as unnecessary loops, redundant computation, or complex branching logic.
Finally, the model jointly considers six primary evidence signals: hotspot coverage, cross-path coverage, within-path repetition, entry proximity, differential self-time, and source-code evidence.
Candidates supported by both source evidence and measured differential cost are preferred.
When the measured evidence is otherwise comparable, candidates are ranked first by the number of differential hotspots they cover, followed by the number of execution paths in which they occur, and then by repeated occurrences within an individual path.
Entry proximity is used to distinguish candidates with similar source and differential evidence.
In addition, the prompt assigns an auxiliary binary \textsc{FastPathPotential} signal when the source code indicates that an upstream controller can introduce a fast path through early return, pruning, or an algorithmic shortcut.
This scheme combines structural coverage, upstream leverage, measured differential cost, and concrete source evidence.
By combining the compact execution paths, the hybrid code context, and these guided reasoning constraints, \textsc{EffiHolmes} enables the LLM to reason about where inefficient logic is introduced and to accurately localize the corresponding fix locations.

\section{Experimental Setup}
\label{Exp_Setup}

% \begin{comment}
\subsection{Research Questions}
\label{Exps: RQs}
In this paper, we focus on the following four research questions. 

\begin{itemize}
    \item \textbf{RQ1 (Effectiveness)}: How effective is EffiHolmes in localizing repository-level inefficiency fix locations?
    \item \textbf{RQ2 (Ablation)}: What are the contributions of the key components in EffiHolmes?
    \item \textbf{RQ3 (Robustness)}: Does \textsc{EffiHolmes} remain robust with smaller models and different reasoning settings?
    % , and can it enable lightweight models to achieve competitive localization performance?
    % llm 的 thinking或者说 reasoning 能力会如何在当前这个任务上影响 EffiHolmes和其他 baseline？
    \item \textbf{RQ4 (Failure Analysis)}: What are the causes of failure cases in \textsc{EffiHolmes}?
    % \item \textbf{RQ5}: What are the characteristics of the cases where EffiHolmes fails, and what insights do they provide for future optimization?
\end{itemize}

\begin{table*}[t]
\centering
\caption{Performance comparison of \textsc{EffiHolmes} and baselines on RepoEffi-Bench. The best results for each model group are highlighted in \textbf{bold} and marked with $^{\star}$. The second-best results are marked with $^{\dagger}$. Relative Gaps between \textsc{EffiHolmes} and the strongest baseline are shown in colour besides \textsc{EffiHolmes'} results.}

\label{tab:main_results}
\resizebox{\textwidth}{!}{%
\begin{tabular}{lllcccccc}
\toprule
\multirow{2}{*}{\textbf{Type}} & \multicolumn{1}{c}{\multirow{2}{*}{\textbf{Method}}} & \multirow{2}{*}{\textbf{Model}} &
\multicolumn{3}{c}{\textbf{File Localization (\%)}} &
\multicolumn{3}{c}{\textbf{Function Localization (\%)}} \\
\cmidrule(lr){4-6} \cmidrule(lr){7-9}
& & & \textbf{Acc@1} & \textbf{Acc@3} & \textbf{Acc@5} & \textbf{Acc@1} & \textbf{Acc@3} & \textbf{Acc@5} \\
\midrule

% --- Retrieval-Based (retained from the original paper; not rerun on current-r2) ---
\multirow{4}{*}{\shortstack{Retrieval-\\Based}}
& E5-base-v2    & - & 24.29 & 39.29 & 42.86 & 3.57  & 11.43 & 12.86 \\
& Jina-Code-v2  & - & 27.86 & 40.00 & 50.00 & 4.29  & 10.00 & 18.57 \\
& CodeSage      & - & 33.57 & 47.14 & 54.29 & 9.29  & 18.57 & 20.00 \\
& CodeRankEmbed & - & 42.14 & 60.00 & 65.00 & 17.86 & 28.57 & 30.71 \\
\midrule

% --- Agent-Based ---
\multirow{15}{*}{\shortstack{Agent-\\Based}}
& \multirow{3}{*}{MoatlessTools}
& \cellcolor{modelA} qwen3-4b
& \cellcolor{modelA} 45.00$^{\dagger}$
& \cellcolor{modelA} 55.00
& \cellcolor{modelA} 60.71
& \cellcolor{modelA} 21.43
& \cellcolor{modelA} 30.00
& \cellcolor{modelA} 35.00 \\
& & \cellcolor{modelB} qwen3-32b
& \cellcolor{modelB} 48.57
& \cellcolor{modelB} 64.29
& \cellcolor{modelB} 66.43
& \cellcolor{modelB} 30.00
& \cellcolor{modelB} 42.86
& \cellcolor{modelB} 47.14 \\
& & \cellcolor{modelC} gpt-5.1
& \cellcolor{modelC} 67.14$^{\dagger}$
& \cellcolor{modelC} 82.14$^{\dagger}$
& \cellcolor{modelC} 85.00
& \cellcolor{modelC} 50.00
& \cellcolor{modelC} 74.29$^{\dagger}$
& \cellcolor{modelC} 78.57$^{\dagger}$ \\
\cmidrule(l){2-9}

& \multirow{3}{*}{SWE-agent}
& \cellcolor{modelA} qwen3-4b
& \cellcolor{modelA} 37.86
& \cellcolor{modelA} 45.00
& \cellcolor{modelA} 47.86
& \cellcolor{modelA} 20.00
& \cellcolor{modelA} 25.00
& \cellcolor{modelA} 26.43 \\
& & \cellcolor{modelB} qwen3-32b
& \cellcolor{modelB} 45.00
& \cellcolor{modelB} 59.29
& \cellcolor{modelB} 62.86
& \cellcolor{modelB} 24.29
& \cellcolor{modelB} 31.43
& \cellcolor{modelB} 31.43 \\
& & \cellcolor{modelC} gpt-5.1
& \cellcolor{modelC} 65.71
& \cellcolor{modelC} 80.00
& \cellcolor{modelC} 85.71
& \cellcolor{modelC} 50.71$^{\dagger}$
& \cellcolor{modelC} 65.71
& \cellcolor{modelC} 72.86 \\
\cmidrule(l){2-9}

& \multirow{3}{*}{OpenHands}
& \cellcolor{modelA} qwen3-4b
& \cellcolor{modelA} 28.57
& \cellcolor{modelA} 38.57
& \cellcolor{modelA} 40.71
& \cellcolor{modelA} 15.00
& \cellcolor{modelA} 20.00
& \cellcolor{modelA} 21.43 \\
& & \cellcolor{modelB} qwen3-32b
& \cellcolor{modelB} 38.57
& \cellcolor{modelB} 53.57
& \cellcolor{modelB} 57.14
& \cellcolor{modelB} 17.86
& \cellcolor{modelB} 22.14
& \cellcolor{modelB} 25.00 \\
& & \cellcolor{modelC} gpt-5.1
& \cellcolor{modelC} 59.29
& \cellcolor{modelC} 75.71
& \cellcolor{modelC} 80.71
& \cellcolor{modelC} 40.00
& \cellcolor{modelC} 60.71
& \cellcolor{modelC} 65.00 \\
\cmidrule(l){2-9}

& \multirow{3}{*}{LocAgent}
& \cellcolor{modelA} qwen3-4b
& \cellcolor{modelA} 42.14
& \cellcolor{modelA} 60.00
& \cellcolor{modelA} 66.43
& \cellcolor{modelA} 25.71$^{\dagger}$
& \cellcolor{modelA} 35.71
& \cellcolor{modelA} 45.00 \\
& & \cellcolor{modelB} qwen3-32b
& \cellcolor{modelB} 49.29
& \cellcolor{modelB} 64.29
& \cellcolor{modelB} 70.00
& \cellcolor{modelB} 32.86$^{\dagger}$
& \cellcolor{modelB} 42.14
& \cellcolor{modelB} 50.71 \\
& & \cellcolor{modelC} gpt-5.1
& \cellcolor{modelC} 66.43
& \cellcolor{modelC} 80.00
& \cellcolor{modelC} 87.14$^{\dagger}$
& \cellcolor{modelC} 48.57
& \cellcolor{modelC} 68.57
& \cellcolor{modelC} 74.29 \\
\cmidrule(l){2-9}

& \multirow{3}{*}{CoSil}
& \cellcolor{modelA} qwen3-4b
& \cellcolor{modelA} 41.43
& \cellcolor{modelA} 50.71
& \cellcolor{modelA} 57.86
& \cellcolor{modelA} 24.29
& \cellcolor{modelA} 32.14
& \cellcolor{modelA} 36.43 \\
& & \cellcolor{modelB} qwen3-32b
& \cellcolor{modelB} 52.86$^{\dagger}$
& \cellcolor{modelB} 63.57
& \cellcolor{modelB} 71.43
& \cellcolor{modelB} 28.57
& \cellcolor{modelB} 43.57
& \cellcolor{modelB} 46.43 \\
& & \cellcolor{modelC} gpt-5.1
& \cellcolor{modelC} 58.57
& \cellcolor{modelC} 67.14
& \cellcolor{modelC} 77.14
& \cellcolor{modelC} 45.00
& \cellcolor{modelC} 57.14
& \cellcolor{modelC} 60.71 \\

\midrule

% --- Procedure-Based ---
\multirow{3}{*}{\shortstack{Procedure-\\Based}}
& \multirow{3}{*}{Agentless}
& \cellcolor{modelA} qwen3-4b
& \cellcolor{modelA} 15.71
& \cellcolor{modelA} 20.00
& \cellcolor{modelA} 21.43
& \cellcolor{modelA} 8.57
& \cellcolor{modelA} 12.86
& \cellcolor{modelA} 15.00 \\
& & \cellcolor{modelB} qwen3-32b
& \cellcolor{modelB} 42.14
& \cellcolor{modelB} 56.43
& \cellcolor{modelB} 67.14
& \cellcolor{modelB} 24.29
& \cellcolor{modelB} 37.14
& \cellcolor{modelB} 45.71 \\
& & \cellcolor{modelC} gpt-5.1
& \cellcolor{modelC} 47.86
& \cellcolor{modelC} 66.43
& \cellcolor{modelC} 75.71
& \cellcolor{modelC} 30.71
& \cellcolor{modelC} 50.00
& \cellcolor{modelC} 55.71 \\
\midrule

% --- Aggregated Profiling-Based ---
\multirow{6}{*}{\shortstack{Profiling-\\Based}}
& \multirow{3}{*}{\shortstack{Direct Aggregated \\ Profiling Prompt}}
& \cellcolor{modelA} qwen3-4b
& \cellcolor{modelA} \textbf{47.86}$^{\star}$
& \cellcolor{modelA} 67.14$^{\dagger}$
& \cellcolor{modelA} 72.86$^{\dagger}$
& \cellcolor{modelA} 22.86
& \cellcolor{modelA} 47.86$^{\dagger}$
& \cellcolor{modelA} 55.71$^{\dagger}$ \\
& & \cellcolor{modelB} qwen3-32b
& \cellcolor{modelB} 52.86$^{\dagger}$
& \cellcolor{modelB} 76.43$^{\dagger}$
& \cellcolor{modelB} 83.57$^{\dagger}$
& \cellcolor{modelB} 27.14
& \cellcolor{modelB} 56.43$^{\dagger}$
& \cellcolor{modelB} 70.00$^{\dagger}$ \\
& & \cellcolor{modelC} gpt-5.1
& \cellcolor{modelC} 63.57
& \cellcolor{modelC} 77.86
& \cellcolor{modelC} 84.29
& \cellcolor{modelC} 48.57
& \cellcolor{modelC} 65.00
& \cellcolor{modelC} 71.43 \\
\cmidrule(l){2-9}

% --- Ours: current-r2; GPT-5.1 uses the official fusion result ---
& \multirow{3}{*}{\textbf{\textsc{EffiHolmes}}}
& \cellcolor{modelA} qwen3-4b
& \cellcolor{modelA} \textbf{47.86}$^{\star}$ {\color{teal}$0.00$}
& \cellcolor{modelA} \textbf{71.43}$^{\star}$ {\color{deepred}$+4.29$}
& \cellcolor{modelA} \textbf{83.57}$^{\star}$ {\color{deepred}$+10.71$}
& \cellcolor{modelA} \textbf{27.14}$^{\star}$ {\color{deepred}$+1.43$}
& \cellcolor{modelA} \textbf{54.29}$^{\star}$ {\color{deepred}$+6.43$}
& \cellcolor{modelA} \textbf{70.71}$^{\star}$ {\color{deepred}$+15.00$} \\

& & \cellcolor{modelB} qwen3-32b
& \cellcolor{modelB} \textbf{60.71}$^{\star}$ {\color{deepred}$+7.85$}
& \cellcolor{modelB} \textbf{81.43}$^{\star}$ {\color{deepred}$+5.00$}
& \cellcolor{modelB} \textbf{87.14}$^{\star}$ {\color{deepred}$+3.57$}
& \cellcolor{modelB} \textbf{40.71}$^{\star}$ {\color{deepred}$+7.85$}
& \cellcolor{modelB} \textbf{64.29}$^{\star}$ {\color{deepred}$+7.86$}
& \cellcolor{modelB} \textbf{75.00}$^{\star}$ {\color{deepred}$+5.00$} \\

& & \cellcolor{modelC} gpt-5.1
& \cellcolor{modelC} \textbf{67.86}$^{\star}$ {\color{deepred}$+0.72$}
& \cellcolor{modelC} \textbf{86.43}$^{\star}$ {\color{deepred}$+4.29$}
& \cellcolor{modelC} \textbf{90.71}$^{\star}$ {\color{deepred}$+3.57$}
& \cellcolor{modelC} \textbf{52.86}$^{\star}$ {\color{deepred}$+2.15$}
& \cellcolor{modelC} \textbf{75.00}$^{\star}$ {\color{deepred}$+0.71$}
& \cellcolor{modelC} \textbf{81.43}$^{\star}$ {\color{deepred}$+2.86$} \\

\bottomrule
\end{tabular}%
}
\end{table*}

\vspace{-1em}

\subsection{Datasets}
We evaluate \textsc{EffiHolmes} on \textsc{RepoEffi-Bench}, a dataset of real-world repository-level inefficiency issues derived from popular Python data science repositories. 
Unlike previous function-level~\cite{shypula2023learning, du2024mercury, huang2024effibench} or synthetic benchmarks~\cite{he2025swe}, RepoEffi-Bench focuses on repository-level time inefficiency issues explicitly reported and fixed by developers.
To ensure practical significance and representativeness, we align repository selection with the quality criteria of SWE-Bench~\cite{jimenez2023swe}.
Starting from the 12 repositories studied in SWE-Bench, we focus on data-science-oriented repositories, since prior work shows that performance bugs are particularly prevalent and impactful in this domain~\cite{yang2025towards}.
We select five repositories that satisfy three requirements: high popularity, active maintenance, and explicit performance-related issue labels.
To identify potential inefficiency reports, we searched for issues using standard keywords established in previous efficiency research~\cite{yang2025towards, jin2012understanding}.
The keywords include: \emph{performance, slow, fast, latency, efficient, optimize, profiling, overhead, timeit}.
We then collect closed inefficiency issues explicitly linked to merged pull requests (PRs), following prior dataset construction practices~\cite{jimenez2023swe,yang2025towards}.
This step yields 1,594 issue--PR pairs.
Following previous work~\cite{chen2025locagent, xia2024agentless}, we filtered these candidates using a two-stage process. LLM-based classification flagged 692 of the 1,594 pairs as potential time-inefficiency issues, and manual verification retained 444 real runtime-slowness cases after removing false positives such as documentation updates. This manual verification did not involve dynamic profiling or scale testing, and the LLM-negative cases were not manually audited.
To ensure reproducibility, we then reconstructed isolated execution environments and retained only cases for which we could reproduce the slowdown on the base commit and verify the speedup on the fix commit using the provided test cases. 
This process further reduced the dataset to 140 instances.

To prevent fix location leakage, we sanitized issue descriptions by removing explicit references to fix locations   (e.g., file paths or function names) while preserving symptom descriptions and reproduction steps. 
The ground truth for localization is derived from the functional code changes in the merged PRs, verified through execution to ensure they directly address the time inefficiency issue. 
The final dataset consists of 140 high-quality instances.

\subsection{Baselines}
\label{Sec3:baseline}
To provide a comprehensive evaluation, we compare EffiHolmes against three categories of competitive baselines, ranging from traditional retrieval techniques to advanced autonomous agents.
\textbf{1) Retrieval-based Methods.}
We evaluate several SOTA embedding models, including the general-purpose \textbf{E5-base-v2} and specialized code embedding models such as \textbf{Jina-Code-v2}, \textbf{Codesage-large-v2}, and the current SOTA model \textbf{CodeRankEmbed}. 
Consistent with standard evaluations, these approaches operate primarily at the function level.
Each function is embedded as a separate unit using a flat indexing structure. 
To preserve semantic context, the function's containing file and class information is appended to its representation before embedding rather than being indexed separately.
\textbf{2) Procedure-based Methods.}
We compare against \textbf{Agentless}~\cite{xia2024agentless}, which employs a structured hierarchical approach to code localization. 
Unlike fully autonomous agents, Agentless utilizes a static repository map to navigate from files to functions without complex iterative planning loops in agent architectures. 
% We use the localization component in Agentless as the baseline.
\textbf{3) Agent-based Methods.}
We include several advanced agent frameworks designed for autonomous code exploration and modification. These baselines consist of \textbf{OpenHands}~\cite{wang2024openhands} (using the default CodeActAgent implementation), \textbf{SWE-Agent}~\cite{yang2024sweagent}, \textbf{Cosil}~\cite{jiang2025cosil}, \textbf{MoatlessTools}~\cite{aorwall_moatless_tools_2025}, which combines search loops with semantic search, and \textbf{LocAgent}~\cite{chen2025locagent}, a graph-guided agent that navigates the codebase through node expansion and relationship traversal. 
These approaches represent the current SOTA in static reasoning and structural navigation, serving as essential baselines.
We adapt these agent-based methods to localization using a common protocol. Each agent receives the target repository, the cleaned issue description, and an instruction to identify the likely fix locations for the reported inefficiency. The agents are restricted to read-only repository inspection and required to return an ordered Top-5 list of file/function candidates. We directly parse this ordered list to compute Acc@1, Acc@3, and Acc@5.

\vspace{-1em}

\subsection{Metrics}
Following previous work~\cite{xia2024agentless, qin2025s, kang2024quantitative, xu2025flexfl, jiang2025cosil}, we use Acc@k (Accuracy at k) to evaluate the performance. 
For each inefficiency issue, the model produces a ranked list of candidate locations. 
We consider a localization attempt successful if the model identifies a relevant fix location (file or function) within the top-$k$ predictions. 
This metric reflects the practical utility of the framework in directing developers toward the correct region for fixing.
We set $k = {1, 3, 5}$, as approximately 73.58\% of developers consider only the top 5 results~\cite{kochhar2016practitioners}.
We report results at multiple granularities: File-Level and Function-Level Localization, each at Acc@1, Acc@3, and Acc@5. 
Importantly, Function-Level localization is only considered correct when the predicted function resides within the correct file. 
Cases where the function matches but the corresponding file does not, are treated as incorrect predictions.

\vspace{-1em}
\subsection{Implementation Details}
Following previous work~\cite{jiang2025cosil}, we employ the Qwen3 models and GPT-5.1 in our experiments.
% 在 rq4 内，为了 研究 reasoning 能力，我们还引入了原生 reasoning 的 qwq-32b
To ensure a fair comparison, we set the temperature to 0.0 as recommended in prior work~\cite{jiang2025cosil}. 
Baseline implementations follow their original papers or official open-source releases~\cite{aorwall_moatless_tools_2025}.
Dynamic profiling is performed using VizTracer, and all experiments run on a Linux server with Ubuntu~20.04, equipped with 128 CPU cores and 8 NVIDIA A6000 GPUs to support concurrent execution and model inference workloads.

\begin{table*}[th]
\centering
\caption{Ablation study results of \textsc{EffiHolmes} on RepoEffi-Bench using GPT-5.1. }
\label{tab:ablation}

\resizebox{\linewidth}{!}{
\begin{tabular}{lcccccc}
\toprule
\multirow{2}{*}{Method} & \multicolumn{3}{c}{File-level Localization (\%)} & \multicolumn{3}{c}{Function-level Localization (\%)} \\
\cmidrule(lr){2-4} \cmidrule(lr){5-7}
 & Acc@1 & Acc@3 & Acc@5 & Acc@1 & Acc@3 & Acc@5 \\
\midrule
\textbf{\textsc{EffiHolmes}}
& \textbf{67.86} & \textbf{86.43} & \textbf{90.71}
& \textbf{52.86} & \textbf{75.00} & \textbf{81.43} \\

w/o Guided Reasoning
& 63.57 {\color{morandi1}$\downarrow$4.29}
& 82.14 {\color{morandi2}$\downarrow$4.29}
& 85.71 {\color{morandi3}$\downarrow$5.00}
& 45.71 {\color{morandi1}$\downarrow$7.14}
& 72.86 {\color{morandi2}$\downarrow$2.14}
& 77.14 {\color{morandi3}$\downarrow$4.29} \\

w/o Execution Paths
& 46.43 {\color{morandi1}$\downarrow$21.43}
& 59.29 {\color{morandi2}$\downarrow$27.14}
& 68.57 {\color{morandi3}$\downarrow$22.14}
& 23.57 {\color{morandi1}$\downarrow$29.29}
& 35.71 {\color{morandi2}$\downarrow$39.29}
& 38.57 {\color{morandi3}$\downarrow$42.86} \\

w/o Execution Paths \& Guided Reasoning
& 44.29 {\color{morandi1}$\downarrow$23.57}
& 57.14 {\color{morandi2}$\downarrow$29.29}
& 65.00 {\color{morandi3}$\downarrow$25.71}
& 22.86 {\color{morandi1}$\downarrow$30.00}
& 34.29 {\color{morandi2}$\downarrow$40.71}
& 37.86 {\color{morandi3}$\downarrow$43.57} \\

\bottomrule
\end{tabular}
}
\end{table*}
\section{Experimental Results}
In this section, we discuss the results of evaluating \textsc{EffiHolmes}  through answering the following four research questions:

\begin{table*}[t]
\centering
\caption{Performance comparison of \textsc{EffiHolmes} and baselines on RepoEffi-Bench (Qwen3-32b).  Annotations indicate the relative improvement of adding reasoning capabilities compared to the baseline (NoReason). The best results for each model group are highlighted in \textbf{bold} and marked with $^{\star}$. The second-best results are marked with $^{\dagger}$. Relative gaps between \textsc{EffiHolmes} and the strongest baseline are shown in colour besides \textsc{EffiHolmes'} results.}
\label{tab:main_results_no_4b}
\resizebox{\textwidth}{!}{%
\begin{tabular}{lllcccccc}
\toprule
\multirow{2}{*}{\textbf{Type}} & \multicolumn{1}{c}{\multirow{2}{*}{\textbf{Method}}} & \multirow{2}{*}{\textbf{Model}} & \multicolumn{3}{c}{\textbf{File Localization (\%)}} & \multicolumn{3}{c}{\textbf{Function Localization (\%)}} \\
\cmidrule(lr){4-6} \cmidrule(lr){7-9}
 & & & \textbf{Acc@1} & \textbf{Acc@3} & \textbf{Acc@5} & \textbf{Acc@1} & \textbf{Acc@3} & \textbf{Acc@5} \\
\midrule

\multirow{8}{*}{\shortstack{Agent-\\Based}}
& \multirow{2}{*}{MoatlessTools}
& \cellcolor{modelA} qwen3-32b (NoReason)
& \cellcolor{modelA} 56.43$^{\dagger}$
& \cellcolor{modelA} 64.29
& \cellcolor{modelA} 66.43
& \cellcolor{modelA} 33.57$^{\dagger}$
& \cellcolor{modelA} 45.71
& \cellcolor{modelA} 47.86 \\
& & \cellcolor{modelC} qwen3-32b
& \cellcolor{modelC} 48.57
& \cellcolor{modelC} 64.29
& \cellcolor{modelC} 66.43
& \cellcolor{modelC} 30.00
& \cellcolor{modelC} 42.86
& \cellcolor{modelC} 47.14 \\

\cmidrule(l){2-9}

& \multirow{2}{*}{SWE-agent}
& \cellcolor{modelA} qwen3-32b (NoReason)
& \cellcolor{modelA} 51.43
& \cellcolor{modelA} 60.71
& \cellcolor{modelA} 66.43
& \cellcolor{modelA} 30.00
& \cellcolor{modelA} 35.00
& \cellcolor{modelA} 38.57 \\
& & \cellcolor{modelC} qwen3-32b
& \cellcolor{modelC} 45.00
& \cellcolor{modelC} 59.29
& \cellcolor{modelC} 62.86
& \cellcolor{modelC} 24.29
& \cellcolor{modelC} 31.43
& \cellcolor{modelC} 31.43 \\

\cmidrule(l){2-9}

& \multirow{2}{*}{OpenHands}
& \cellcolor{modelA} qwen3-32b (NoReason)
& \cellcolor{modelA} 45.71
& \cellcolor{modelA} 62.14
& \cellcolor{modelA} 65.00
& \cellcolor{modelA} 20.71
& \cellcolor{modelA} 25.00
& \cellcolor{modelA} 26.43 \\
& & \cellcolor{modelC} qwen3-32b
& \cellcolor{modelC} 38.57
& \cellcolor{modelC} 53.57
& \cellcolor{modelC} 57.14
& \cellcolor{modelC} 17.86
& \cellcolor{modelC} 22.14
& \cellcolor{modelC} 25.00 \\

\cmidrule(l){2-9}

& \multirow{2}{*}{LocAgent}
& \cellcolor{modelA} qwen3-32b (NoReason)
& \cellcolor{modelA} 51.43
& \cellcolor{modelA} 71.43
& \cellcolor{modelA} 73.57
& \cellcolor{modelA} 31.43
& \cellcolor{modelA} 50.00
& \cellcolor{modelA} 52.14 \\
& & \cellcolor{modelC} qwen3-32b
& \cellcolor{modelC} 49.29
& \cellcolor{modelC} 64.29
& \cellcolor{modelC} 70.00
& \cellcolor{modelC} 32.86$^{\dagger}$
& \cellcolor{modelC} 42.14
& \cellcolor{modelC} 50.71 \\

\midrule

\multirow{4}{*}{\shortstack{Procedure-\\Based}}
& \multirow{2}{*}{CoSIL}
& \cellcolor{modelA} qwen3-32b (NoReason)
& \cellcolor{modelA} 50.00
& \cellcolor{modelA} 57.86
& \cellcolor{modelA} 66.43
& \cellcolor{modelA} 29.29
& \cellcolor{modelA} 42.14
& \cellcolor{modelA} 47.14 \\
& & \cellcolor{modelC} qwen3-32b
& \cellcolor{modelC} 52.86$^{\dagger}$
& \cellcolor{modelC} 63.57
& \cellcolor{modelC} 71.43
& \cellcolor{modelC} 28.57
& \cellcolor{modelC} 43.57
& \cellcolor{modelC} 46.43 \\

\cmidrule(l){2-9}

& \multirow{2}{*}{Agentless}
& \cellcolor{modelA} qwen3-32b (NoReason)
& \cellcolor{modelA} 36.43
& \cellcolor{modelA} 58.57
& \cellcolor{modelA} 66.43
& \cellcolor{modelA} 19.29
& \cellcolor{modelA} 35.00
& \cellcolor{modelA} 41.43 \\
& & \cellcolor{modelC} qwen3-32b
& \cellcolor{modelC} 42.14
& \cellcolor{modelC} 56.43
& \cellcolor{modelC} 67.14
& \cellcolor{modelC} 24.29
& \cellcolor{modelC} 37.14
& \cellcolor{modelC} 45.71 \\

\midrule

\multirow{4}{*}{\shortstack{Profiling-\\Based}}
& \multirow{2}{*}{\shortstack{Direct Aggregated \\ Profiling Prompt}}
& \cellcolor{modelA} qwen3-32b (NoReason)
& \cellcolor{modelA} 55.00
& \cellcolor{modelA} 76.43$^{\dagger}$
& \cellcolor{modelA} \textbf{86.43}$^{\star}$
& \cellcolor{modelA} 28.57
& \cellcolor{modelA} 53.57$^{\dagger}$
& \cellcolor{modelA} 70.71$^{\dagger}$ \\
& & \cellcolor{modelC} qwen3-32b
& \cellcolor{modelC} 52.86$^{\dagger}$
& \cellcolor{modelC} 76.43$^{\dagger}$
& \cellcolor{modelC} 83.57$^{\dagger}$
& \cellcolor{modelC} 27.14
& \cellcolor{modelC} 56.43$^{\dagger}$
& \cellcolor{modelC} 70.00$^{\dagger}$ \\

\cmidrule(l){2-9}

& \multirow{2}{*}{\textbf{\textsc{EffiHolmes}}}
& \cellcolor{modelA} qwen3-32b (NoReason)
& \cellcolor{modelA} \textbf{61.43}$^{\star}$ {\color{deepred}$+5.00$}
& \cellcolor{modelA} \textbf{77.14}$^{\star}$ {\color{deepred}$+0.71$}
& \cellcolor{modelA} 85.00$^{\dagger}$ {\color{teal}$-1.43$}
& \cellcolor{modelA} \textbf{38.57}$^{\star}$ {\color{deepred}$+5.00$}
& \cellcolor{modelA} \textbf{60.00}$^{\star}$ {\color{deepred}$+6.43$}
& \cellcolor{modelA} \textbf{71.43}$^{\star}$ {\color{deepred}$+0.72$} \\
& & \cellcolor{modelC} qwen3-32b
& \cellcolor{modelC} \textbf{60.71}$^{\star}$ {\color{deepred}$+7.86$}
& \cellcolor{modelC} \textbf{81.43}$^{\star}$ {\color{deepred}$+5.00$}
& \cellcolor{modelC} \textbf{87.14}$^{\star}$ {\color{deepred}$+3.57$}
& \cellcolor{modelC} \textbf{40.71}$^{\star}$ {\color{deepred}$+7.86$}
& \cellcolor{modelC} \textbf{64.29}$^{\star}$ {\color{deepred}$+7.86$}
& \cellcolor{modelC} \textbf{75.00}$^{\star}$ {\color{deepred}$+5.00$} \\

\bottomrule
\end{tabular}
}
\end{table*}

\subsection{RQ1. Effectiveness}
\label{sec:rq1}
\noindent\textit{Question.} How effective is \textsc{EffiHolmes} in localizing repository-level inefficiency fix locations compared with existing methods?

\noindent\textit{Approach.} We compare \textsc{EffiHolmes} with representative retrieval-based, procedure-based, and agent-based baselines on RepoEffi-Bench.
We also include a \emph{Direct Aggregated Profiling} baseline that provides only per-function runtime statistics without execution-path structure.
This comparison isolates the impact of runtime execution paths with calling relationships and assesses whether \textsc{EffiHolmes} offers significant advantages beyond static similarity matching, iterative exploration, or aggregated profiling heuristics.

\noindent\textit{Results.} \textbf{\textsc{EffiHolmes} achieves better performance than the baselines and the direct aggregated profiling baseline on nearly all evaluation metrics.} 
% As shown in Table~\ref{tab:main_results}, \textsc{EffiHolmes} achieves the best results in 17 out of 18 evaluation scenarios (6 metrics $\times$ 3 models).
As shown in Table~\ref{tab:main_results}, \textsc{EffiHolmes} achieves the best or tied-best results in all 18 evaluation scenarios (6 metrics $\times$ 3 models), including 17 unique best results and one tied-best result.
% , accounting for 94.4\% of all settings. 
On GPT-5.1, it reaches a file-level Acc@5 of 90.71\%, outperforming the strongest baseline, LocAgent, by 3.57 pp.
This result indicates that \textsc{EffiHolmes} can reliably identify the correct file containing the inefficiency fix within the top-5 candidates, even in large and complex repositories.

\textbf{\textsc{EffiHolmes} shows its clearest advantage on the smaller model at higher values of $k$.}
On qwen3-4b, the performance gaps between \textsc{EffiHolmes} and the strongest baselines are modest at Acc@1 but become larger at Acc@3 and Acc@5.
In particular, \textsc{EffiHolmes} exceeds the strongest baseline, Direct Aggregated Profiling, by 15.00 pp on function-level Acc@5.
This suggests that the differential profiling-guided execution paths are particularly helpful under limited model capacity, as they provide structured runtime evidence that supports more effective candidate ranking at broader localization cutoffs.

\textbf{However, the advantages of \textsc{EffiHolmes} are not uniformly large at low localization cutoffs.}
On qwen3-4b, \textsc{EffiHolmes} ties Direct Aggregated Profiling on file-level Acc@1 at 47.86\%.
On GPT-5.1, its official-fusion result reaches 67.86\% file-level Acc@1, exceeding the agent-based MoatlessTools baseline at 67.14\% by only 0.71 pp.
Direct profiling and agent-based methods can remain competitive at Top-1 when the issue description contains explicit API-level clues that map directly to a specific file or function.
For example, in pandas \#43549, the issue description repeatedly names \texttt{MultiIndex.equals}, which maps almost directly to \texttt{pandas/core/indexes/multi.py::equals}.
Consequently, Direct Aggregated Profiling on qwen3-4b and MoatlessTools on GPT-5.1 both rank this ground-truth location first, whereas \textsc{EffiHolmes} ranks it fifth and second, respectively, after prioritizing lower-level callees exposed by execution evidence.
Even in such cases, \textsc{EffiHolmes} still retrieves the correct location within the Top-5 candidates on qwen3-4b and within the Top-2 candidates on GPT-5.1, while retaining clearer aggregate advantages at larger values of $k$ and at function-level localization.

\textbf{The limitations of existing baselines are more evident at function-level localization.}
Retrieval-based methods perform poorly in this setting: even the best retriever, CodeRankEmbed, reaches only 17.86\% function-level Acc@1, suggesting that inefficiency fix locations cannot be reliably identified from text similarity or static code structure alone.
Agent-based frameworks are competitive on large models, but much less robust on smaller ones: among the complete runs, their function-level Acc@1 ranges from 40.00\% to 50.00\% on GPT-5.1, but decreases to 15.00\%--25.71\% on qwen3-4b.
This suggests that multi-step exploration depends strongly on model capacity.
Direct Aggregated Profiling further shows that runtime profiling is useful, but insufficient by itself: although it reaches 48.57\% function-level Acc@1 and 71.43\% Acc@5 on GPT-5.1, aggregated profiling flattens call-stack structure and lacks execution-path context, making it harder to reason about fix locations.

\subsection{RQ2. Ablation}
\label{sec:rq2}
\noindent\textit{Question.} What are the contributions of the key components in \textsc{EffiHolmes}?

\noindent\textit{Approach.} We perform an ablation study on GPT-5.1 by removing \emph{Critical Execution Paths Extraction}, \emph{Guided Reasoning}, or both, while keeping the rest of the pipeline unchanged. This design isolates the individual contribution of structured execution-path evidence and reasoning constraints, as well as their combined effect in the full framework.
We define the ablation variants as follows. 
\textbf{w/o Execution Paths} removes the filtered critical execution paths from the guided-reasoning prompt, while retaining the rest. 
\textbf{w/o Guided Reasoning} keeps the paths, issue descriptions and code contexts, but removes the explicit reasoning and ranking instructions from the prompt.
\textbf{w/o Execution Paths \& Guided Reasoning} removes both the paths and the guided reasoning constraints, leaving the model to predict fix locations from the remaining issue description and code context alone.

\noindent\textit{Results.} 
\textbf{Both Critical Execution Paths Extraction and Guided Reasoning contribute to the effectiveness of \textsc{EffiHolmes}, with execution paths providing the larger contribution and the two components showing complementary effects.}
As shown in Table~\ref{tab:ablation}, the full model consistently outperforms all ablated variants across all evaluation metrics.
Removing either component causes degradation, while removing both leads to the largest drop across all six metrics, with function-level Acc@1 falling to 22.86\%, indicating that the two components play complementary roles in the full framework.

\textbf{Critical Execution Paths Extraction is the primary source of improvement.}
Removing execution paths causes the largest degradation among all single-component ablations.
Compared with the full model, \textsc{EffiHolmes} without execution paths drops by 39.29 pp on function-level Acc@3 and by 21.43 pp on file-level Acc@1.
This shows that compact execution paths provide the key structured runtime clues for narrowing the search space and linking hotspots to upstream fix locations.

\textbf{Removing the guided reasoning rules leads to a consistent but comparatively smaller decline in performance.}
Although its effect is smaller than that of execution paths, removing guided reasoning still causes a consistent decline across metrics, including 7.14 pp on function-level Acc@1 and 4.29 pp on file-level Acc@3.
These results confirm that the guided reasoning rules contribute meaningfully to the overall effectiveness of our method.

\vspace{-1em}

\begin{table*}[t]
\centering
\caption{Performance comparison of Profiling-Based methods on RepoEffi-Bench (qwen3-4b). Annotations indicate the relative improvement of adding reasoning capabilities (\textit{qwen3-4b}) compared to the baseline (\textit{NoReason}).}
\label{tab:profiling_comparison_4b}
\resizebox{\textwidth}{!}{%
\begin{tabular}{lllcccccc}
\toprule
\multirow{2}{*}{\textbf{Type}} & \multicolumn{1}{c}{\multirow{2}{*}{\textbf{Method}}} & \multirow{2}{*}{\textbf{Model}} & \multicolumn{3}{c}{\textbf{File Localization (\%)}} & \multicolumn{3}{c}{\textbf{Function Localization (\%)}} \\
\cmidrule(lr){4-6} \cmidrule(lr){7-9}
 & & & \textbf{Acc@1} & \textbf{Acc@3} & \textbf{Acc@5} & \textbf{Acc@1} & \textbf{Acc@3} & \textbf{Acc@5} \\
\midrule

\multirow{4}{*}{\shortstack{Profiling-\\Based}}
& \multirow{2}{*}{\shortstack{Direct Aggregated \\ Profiling Prompt}}
& qwen3-4b (NoReason)
& 55.00 & 68.57 & 75.71 & 25.71 & 44.29 & 55.00 \\

&
& qwen3-4b
& 47.86 {\color{teal}$\downarrow$7.14}
& 67.14 {\color{teal}$\downarrow$1.43}
& 72.86 {\color{teal}$\downarrow$2.86}
& 22.86 {\color{teal}$\downarrow$2.86}
& 47.86 {\color{deepred}$\uparrow$3.57}
& 55.71 {\color{deepred}$\uparrow$0.71} \\

\cmidrule(l){2-9}

& \multirow{2}{*}{\textbf{\textsc{EffiHolmes}}}
& qwen3-4b (NoReason)
& 53.57 & 75.71 & 85.00 & 28.57 & 57.86 & 72.14 \\

&
& qwen3-4b
& 47.86 {\color{teal}$\downarrow$5.71}
& 71.43 {\color{teal}$\downarrow$4.29}
& 83.57 {\color{teal}$\downarrow$1.43}
& 27.14 {\color{teal}$\downarrow$1.43}
& 54.29 {\color{teal}$\downarrow$3.57}
& 70.71 {\color{teal}$\downarrow$1.43} \\

\bottomrule
\end{tabular}
}
\end{table*}

\subsection{RQ3. Robustness}
\label{sec:rq3}
\noindent\textit{Question.} Does \textsc{EffiHolmes} remain robust with smaller models and different reasoning settings?

\noindent\textit{Approach.} We evaluate robustness along two axes: (\romannumeral1) model scaling, from GPT-5.1 and qwen3-32b to qwen3-4b; and (\romannumeral2) inference-time reasoning toggling under model scaling. For 4B-scale models, we restrict the comparison to non-agent methods, because preliminary runs showed that agent workflows often fail to complete long-horizon multi-step exploration reliably at this scale, which would confound the analysis of reasoning and context structure.

\noindent\textit{Results.} 
\textbf{\textsc{EffiHolmes} remains substantially more robust than the baselines at the Top-5 as model capacity decreases.}
As shown in Table~\ref{tab:main_results}, when the model shifts from GPT-5.1 to qwen3-4b, \textsc{EffiHolmes} exhibits the smallest performance degradation at the Top-5 among the compared methods.
At the function level, its Acc@5 drops by only 10.71 pp, whereas Direct Aggregated Profiling and MoatlessTools drop by 15.71 and 43.57 pp, respectively.
A similar pattern holds at the file level: \textsc{EffiHolmes} decreases by 7.14 pp, compared with 11.43 pp for Direct Aggregated Profiling and 24.29 pp for MoatlessTools.
These results indicate that the execution-path clues produced by \textsc{EffiHolmes} remain effective even for lightweight models at higher values of $k$, while baseline methods degrade more sharply under reduced model capacity.

\textbf{\textsc{EffiHolmes} also enables lightweight models to rival, and in some cases surpass, larger-model baselines.}
The qwen3-4b+\textsc{EffiHolmes} configuration achieves a function-level Acc@5 that is 0.71 pp higher than the best qwen3-32b baseline (70.00\% for Direct Aggregated Profiling), and its file-level Acc@5 matches the best qwen3-32b result at 83.57\%.
Even under stricter settings, its low-$k$ performance can exceed the high-$k$ performance of larger baselines; for example, its function-level Acc@3 of 54.29\% already surpasses the function-level Acc@5 of several qwen3-32b baselines.
This suggests that profiling-guided execution paths can partially compensate for smaller model size by providing more explicit structured runtime evidence for repository-level localization.

\textbf{Robustness does not consistently hold across inference-time reasoning settings under model scaling.}
On qwen3-32b (Table~\ref{tab:main_results_no_4b}), enabling reasoning improves \textsc{EffiHolmes} on most metrics at both file and function levels.
For example, function-level Acc@1 increases from 38.57\% to 40.71\%, although file-level Acc@1 slightly decreases from 61.43\% to 60.71\%.
More importantly, this gain does not persist on qwen3-4b (Table~\ref{tab:profiling_comparison_4b}): file-level Acc@1 decreases from 53.57\% to 47.86\% and function-level Acc@5 decreases from 72.14\% to 70.71\%.
On qwen3-4b, enabling reasoning also reduces the file-level performance of Direct Aggregated Profiling, where File Acc@1 drops from 55.00\% to 47.86\% and File Acc@3 falls from 68.57\% to 67.14\%, although its function-level Acc@3 and Acc@5 improve slightly.
Overall, these results suggest that the effect of inference-time reasoning is model- and method-dependent, rather than uniformly beneficial under limited model capacity.

\subsection{RQ4. Failure Analysis}
\label{sec:rq4}
\noindent\textit{Question.} What are the causes of failure cases in \textsc{EffiHolmes}?

\noindent\textit{Approach.} We manually inspect the cases where \textsc{EffiHolmes} with GPT-5.1 fails to localize the ground-truth fix within function-level Acc@5. 
We focus on this setting because function-level Top-5 is primary
developer-facing target: it represents a practical debugging shortlist while remaining strict enough that misses are still meaningful to inspect~\cite{kochhar2016practitioners}. 
We conduct this analysis with GPT-5.1 to factor out model-capacity effects and focus on method-level limitations.

\noindent\textit{Results.} 
We find that these failures primarily occur when the ground-truth fix lies outside the Python-level localization scope (e.g., implemented in native C/Cython code), or when the profiler’s time resolution is too coarse to capture low-cost functions, causing them to be absent from the extracted execution paths. Based on our analysis, we identify two dominant failure modes:

\textbf{First, some failures are caused by scope mismatch beyond Python-level localization.} \textsc{EffiHolmes} currently localizes fix sites within Python functions represented in the extracted Python-level execution paths.
However, for certain inefficiency issues, the ground-truth fix is implemented in native code (e.g., C/Cython) that is invoked by Python wrappers, which is outside our current localization scope.
For example, in pandas \#5765, the fix is implemented in native C functions under \texttt{pandas/src/ujson/}, including \texttt{JSON\_EncodeObject} and the \texttt{NpyArr\_*} encoding helpers.
In contrast, the extracted paths primarily expose Python-level callers such as \texttt{pandas/io/json.py::to\_json}.
Consequently, \textsc{EffiHolmes} ranks these visible wrapper functions but cannot reach the native ground-truth locations.

\textbf{Second, some fixes fall outside the extracted execution paths due to profiling coverage limitations.}
In \texttt{pandas} \#44106 case, the ground-truth function \path{CParserWrapper._set_noconvert_columns} in \path{pandas/io/parsers/c_parser_wrapper.py} does not appear in the extracted execution paths.
Further inspection of the raw profiling traces also finds no occurrence of this function, indicating limited visibility of this thin wrapper.
Before the fix, the function maps parser column names to their positions by repeatedly calling \texttt{self.orig\_names.index(x)}.
These repeated linear searches cause quadratic behavior, but their runtime cost is attributed primarily to the built-in \texttt{list.index}, leaving the wrapper itself with little visible profiling signal.

\section{Related Works}
% In this section, we review prior work closely related to our research, focusing on fault localization and code optimization.
\subsection{Fault Localization}

\subsubsection{Traditional Fault Localization}
Traditional FL techniques are largely built upon explicit correctness signals. 
\textit{SBFL} methods~\cite{jones2005tarantula,abreu2007ochiai,abreu2009spectrum,raselimo2019spectrumCFG,reis2019demystifying,wong2014dstar,zhang2011localizing} rank suspicious code by statistically contrasting execution coverage between passing and failing tests. 
\textit{MBFL} approaches, such as Metallaxis and MUSE~\cite{papadakis2013metallaxis,moon2014muse}, inject artificial mutations and infer fault relevance from the resulting test outcomes. 
\textit{MLFL} methods, including DEEPRL4FL~\cite{li2021fault}, DeepFL~\cite{li2019deepfl}, and GRACE~\cite{lou2021boosting}, further learn predictive models over such coverage- and mutation-derived features. 
Despite their differences, they rely on a key assumption: the availability of a binary oracle that distinguishes failing from passing executions. 
This assumption does not hold for repository-level inefficiency issues, which manifest as performance degradation without producing failing executions~\cite{jin2012understanding,han2016empirical}.
Consequently, these methods are not applicable to our setting.

\subsubsection{LLM-Based Fault Localization}
Recent studies have shown that LLMs are highly effective for fault localization, especially in issue-driven and repository-level settings. 
Representative methods include direct or hierarchical prompting approaches such as Agentless~\cite{xia2024agentless}, structure-aware localization methods such as OrcaLoca~\cite{yu2025orcaloca}, LocAgent~\cite{chen2025locagent}, and COSIL~\cite{jiang2025cosil}, as well as agent-based systems such as SWE-Agent~\cite{yang2024sweagent}, AutoCodeRover~\cite{zhang2024autocoderover}, OpenHands CodeAct~\cite{wang2024openhands}, Moatless Tools~\cite{aorwall_moatless_tools_2025}, and SWE-Search~\cite{antoniades2024swe}. 
These methods differ in search strategy and tool support, but they share a critical underlying assumption: they rely heavily on explicit localization clues from issue descriptions or failure artifacts, such as stack traces, error messages, and failing tests, to narrow the search space in large repositories. 
However, this assumption does not hold for time inefficiency issues, which rarely provide such explicit clues and instead manifest as performance degradation without observable failures. Therefore, directly reusing existing LLM-based FL pipelines is ineffective for repository-level inefficiency localization. To address this gap, we adopt runtime profiling as the primary clue source, and make it usable for this setting through differential signal amplification, critical-path extraction, and guided upstream reasoning over structured runtime evidence.

\subsection{Code Optimization}
Early work on code optimization primarily relied on rule-based and program-analysis techniques to identify specific inefficiency patterns, such as software misconfigurations, redundant computations, and loop inefficiencies~\cite{nistor2013discovering,iqbal2021cadet,della2015performance,giavrimis2021genetic}. Representative examples include \textsc{MemoizeIt}, which uses dynamic analysis to detect redundant computations~\cite{della2015performance}, and \textsc{CLARITY}, which applies static analysis to identify redundant traversal bugs~\cite{olivo2015static}. These methods are effective for known optimization patterns, but typically depend on expert-crafted rules or specialized analyses.

With the advent of LLMs, research has shifted toward generative and agentic optimization frameworks. Some methods directly synthesize optimized patches using pretrained models, such as \textsc{RAPGen}~\cite{garg2025rapgen}, \textsc{DeepPERF}~\cite{garg2022deepdev}, and \textsc{Supersonic}~\cite{chen2024supersonic}, while benchmarks such as PIE~\cite{shypula2023learning} evaluate prompting strategies for code speedup. More recent work adopts iterative, search-based, or multi-agent workflows, including \textsc{MARCO}~\cite{rahman2025marco}, \textsc{Artemis AI}~\cite{giavrimis2025artemis}, \textsc{SemOpt}~\cite{zhao2025semopt}, MPCO~\cite{gong2025tuning}, and human--LLM collaborative optimization~\cite{florath2023llm}.

However, most existing optimization methods operate at the function level: they assume that the inefficient code region or hotspot has already been identified, and focus on \emph{how} to optimize it. In contrast, \textsc{EffiHolmes} addresses the earlier and orthogonal problem of \emph{where} to fix a repository-level time inefficiency. By integrating differential profiling with domain-guided reasoning, our approach localizes fix locations within large repositories instead of rewriting a pre-selected function.

% \vspace{-0.5em}
\section{Threats to Validity}

We discuss the threats to validity in this section.
One threat is that, in this paper, we conduct experiments using GPT-5.1's APIs. 
Since the model is closed-source, its training data are not publicly accessible, which raises concerns about potential data leakage. 
However, this threat is partially mitigated by our empirical results: GPT-5.1 does not achieve strong performance when used directly in the baseline approaches. 
This suggests that the fix locations identified by \textsc{EffiHolmes} are unlikely to result merely from memorization of training data, but instead stem from the profiling-guided execution paths and reasoning mechanisms introduced by our approach.

Another threat is that our evaluation focuses only on Python repositories and Python-level profiling. 
As a result, the current study does not directly cover cases where the true fix is implemented in native C/Cython code behind Python wrappers, or cases where the true fix location is absent from the extracted execution paths due to weak profiling visibility.
Nevertheless, the core ideas of differential profiling and execution-path--guided reasoning are not inherently specific to Python. 
Extending \textsc{EffiHolmes} to other programming languages and runtime platforms is therefore plausible, although its effectiveness beyond the current scope still requires future validation.
In addition, an off-path fix location cannot be recovered by the current pipeline, because the guided reasoning stage ranks candidates drawn from these paths. 
This limitation affects the method's maximum achievable recall, although such instances remain included in the evaluation denominator.

% \vspace{-1em}
\section{Conclusion}
% 总领句
We propose \textsc{EffiHolmes}, an LLM-based framework for localizing fix locations of repository-level time inefficiency issues. 
% 做了啥，为啥做
EffiHolmes employs differential profiling to pinpoint inefficiency hotspots effectively.
It performs execution path distillation to disentangle execution paths relevant to inefficiency, constructing compact paths that connect hotspots to the reported inefficient function. 
Guided LLM reasoning with domain heuristics is introduced to bridge the semantic gap between the observed inefficiencies and the actual fix locations. 
In addition, we introduce \textsc{RepoEffi-Bench}, the first benchmark specifically designed for repository-level inefficiency localization. 
% 效果咋样
Experiments on \textsc{RepoEffi-Bench} show that \textsc{EffiHolmes} outperforms the best baselines by 4.29 pp on GPT-5.1 file-level Acc@3 and 15.00 pp on qwen3-4b function-level Acc@5.
In addition, it remains robust across different model capabilities.

\section{Data Availability}
The code and dataset is available at \url{https://github.com/HaowenYoung/ASE26_EffiHolmes}.

\bibliographystyle{ACM-Reference-Format}
\balance
\bibliography{FSE_ref_corrected}

\end{document}